\documentstyle[12pt,epsf]{article}
\def\d  {\delta}

\def\l  {\lambda}

\def\o  {\omega}

\def\p  {\pi}
\def\P  {\Pi}

\newcommand{\ba}{\begin{array}}
\newcommand{\ea}{\end{array}}
\newcommand{\beq}{\begin{equation}}
\newcommand{\eeq}{\end{equation}}
\newcommand{\bea}{\begin{eqnarray}}
\newcommand{\eea}{\end{eqnarray}}
\newcommand{\beal}{\setcounter{letter}{1} \begin{eqnarray}}
\newcommand{\eeal}{\addtocounter{equation}{1} \end{eqnarray}}

\newcommand{\req}[1]{Eq.(\ref{#1})}

\newcommand{\larrow}{\,\,\,\,\hbox to 30pt{\rightarrowfill}
\,\,\,\,}
\newcommand{\slarrow}{\,\,\,\hbox to 20pt{\rightarrowfill}
\,\,\,}

\newcommand{\IR}{{\rm I\kern-.22em R}}
\begin{document}

\begin{titlepage}
\renewcommand{\thefootnote}{\fnsymbol{footnote}}
\renewcommand{\baselinestretch}{1.3}
\medskip

\begin{center}
{\large {\bf Spectrum of Charged Black Holes -- The Big Fix Mechanism
Revisited\\ }}
\medskip

\renewcommand{\baselinestretch}{1}
{\bf
Andrei Barvinsky $\dagger$
Saurya Das $\sharp$
Gabor Kunstatter $\sharp$
\\}
\vspace*{0.50cm}
{\sl
$\dagger$ Theory Department\\
Lebedev Physics Institute and Lebedev Research Center in Physics\\
Leninsky Prospect 53, Moscow 117924, Russia\\
{[e-mail: barvin@td.lpi.ru]}\\ [5pt]
}
{\sl
$\sharp$ Dept. of Physics and Winnipeg Institute for
Theoretical Physics, University of Winnipeg\\
Winnipeg, Manitoba, Canada R3B 2E9\\
{[e-mail: gabor@theory.uwinnipeg.ca, saurya@theory.uwinnipeg.ca]}\\[5pt]
 }

\end{center}

\renewcommand{\baselinestretch}{1}

\begin{center}
{\bf Abstract}
 \end{center}
{\small Following an earlier suggestion of the
authors (gr-qc/9607030), we use some basic properties of Euclidean
black hole thermodynamics and the quantum mechanics of systems
with periodic phase space coordinate to derive the discrete
two-parameter area spectrum of generic charged spherically
symmetric black holes in any dimension. For the Reissner-Nordstrom
black hole we get $A/4G\hbar=\pi(2n+p+1)$, where the integer
$p=0,1,2,..$ gives the charge spectrum, with $Q=\pm\sqrt{\hbar
p}$. The quantity $\pi(2n+1)$, $n=0,1,...$ gives a measure of the
excess of the mass/energy  over the critical minimum (i.e.
extremal) value allowed for a given fixed charge $Q$. The
classical critical bound cannot be saturated due to vacuum
fluctuations of the horizon, so that generically extremal black
holes do not appear in the physical spectrum. Consistency also
requires the black hole charge to be an integer multiple of any
fundamental  elementary particle charge: $Q=\pm me$,
$m=0,1,2,...$. As a by-product this yields a relation between the
fine structure constant and integer parameters of the black hole
-- a kind of the Coleman big fix mechanism induced by black
holes. In four dimensions, this relationship is $e^2/\hbar=p/m^2$
and requires the fine structure constant to be a rational number.
Finally, we prove that the horizon area is an adiabatic
invariant, as has been conjectured previously.} \vfill \hfill


\end{titlepage}
\clearpage
\section{Introduction}

One of the most important unsolved problems in theoretical physics
concerns the synthesis of quantum mechanics and  general
relativity. By the very  nature of the problem, it is difficult,
if not impossible, to find experimental clues as to what form such
a synthesis might take. Candidate theories like string theory and
quantum geometry do not as yet have experimentally verifiable
predictions. It is therefore useful to examine theoretical
arguments about what to expect generically from a quantum theory
of gravity. In this context, black holes provide an ideal
theoretical laboratory.

There is a great deal of evidence for the existence of black holes
in binary systems and at the center of many galaxies, including
our own. The work of Bekenstein and Hawking in the mid-seventies
has shown that black holes behave as thermodynamic systems, where
the surface gravity and horizon area represent the temperature and
entropy, respectively. Moreover, black holes emit thermal
radiation via quantum processes near the horizon.  The
Bekenstein-Hawking \cite{BH} entropy $S_{BH}$ for a black hole is
proportional to the area $A$  of its (outer) horizon:
    \bea S_{BH}=
    {A\over 4 G\hbar} \label{bh entropy}\,.
    \eea
This formula is assumed to be generically valid for black holes in
any spacetime dimension. As a specific example, in spherically
symmetric Einstein-Maxwell theory in 4 dimensions we have  the
Reissner-Nordstr\"om black hole. In this case (we work in units
where $c=1$):
    \bea A=4\pi \left(GM+
    \sqrt{G^2M^2-GQ^2}\right)^2.  \label{rn entropy}
    \eea

The microscopic origin of this thermodynamic behavior is yet to be
understood in general, and it is commonly believed that such an
understanding can only be achieved in the context of a quantum
theory of gravity. One fundamental  question that has been asked
in recent years is: what is the quantum spectrum of the
fundamental observables, namely mass and charge? The answer to
this question will determine the transition rates between quantum
states and hence will have observable consequences for the Hawking
radiation spectrum.  Bekenstein and Mukhanov \cite{bm0,bm} have
argued from very general grounds that the area of quantum black
holes (and hence the entropy) should have a uniformly spaced
spectrum, of the form :
    \bea
    A \propto n, \qquad n=0,1,2,...\,. \label{area quantization}
    \eea
Bekenstein's arguments are based in part on a conjectured
relationship between horizon area and adiabatic invariants
\cite{bm}, which by the Bohr-Sommerfeld quantization rule, always
have a discrete spectrum. In Bekenstein's own words, these
arguments involve ``a mixture of classical hints and quantum
ideas''.

The purpose of this paper is to derive a precise form of
\req{area quantization} from essentially one important assumption
which  encodes the semi-classical thermodynamic content of black
hole dynamics, following the analysis developed in \cite{bk}. In
particular, we will assume that $P_M$, the variable conjugate to
the black hole mass $M$ is periodic, with period equal to the
inverse Hawking temperature associated with the black hole. This
is strictly true only in the Euclidean (imaginary time) sector of
the theory. However, this single assumption, plus a natural
periodicity condition on the $U(1)$ phase associated with the
electromagnetic potential allow us to derive rigorously both the
area and charge quantization conditions from standard quantum
mechanics. Other derivations of spectra similar to \req{area
quantization} exist, but these are tied to specific models
\cite{Berezin}, in particular theories of gravity \cite{others} and
periodicity assumptions in {\em Lorentzian} time (imposed by hand
in \cite{kastrup} or attempted to be justified in \cite{louko} on
account of bounded motion of the Einstein-Rosen bridge throat in
Kruskal spacetime). To the best of our knowledge, though, no
previous work has utilized the periodicity of the $U(1)$ phase to
obtain a charge quantization condition  in addition to  the area
spectrum. Moreover, it is the interplay between the  periodicity
of the imaginary time and that of the $U(1)$ phase that ultimately
yields the interesting constraint on the fine structure constant.
One important consequence of our analysis is that near extremal black
holes appear in the spectrum as highly quantum (low quantum number)
objects.
Another direct outcome is a proof of
Bekenstein's conjecture that area (and entropy) is an adiabatic
invariant associated with black hole dynamics.

It is important to note that our analysis is
very general: it is valid for a large class of charged, or uncharged black
holes. This class includes spherically symmetric black holes in Einstein
gravity (with
or
without cosmological constant) in any dimension as well as the
rotating BTZ black hole, with the modification that the ``charge'' actually
corresponds to the angular momentum of the black hole.

The paper is organized as follows: Section 2 defines what we mean by
``generic black holes'', and reviews their dynamical and  thermodynamic
properties in the context
of generic 2-D dilaton graviton. Section 3 presents the quantization
of two well known toy models in order to motivate the methodology that
we apply in Section 4 to derive the charge and area spectrum
 of generic black holes. In Section 5, some consequences of our
quantization are derived, while Section 6 contains a summary and
conclusions.

\section{Generic Charged Black Holes}

Consider a spherically symmetric metric in $d$ spacetime
dimensions of the form: \bea ds^2= g_{\alpha\beta}dx^\alpha
dx^\beta + r^2(x,t)d\Omega^{(d-2)}\, , \label{metric1} \eea where
$x^\alpha$ denotes the coordinates of the radial and time parts of
the metric, while, $d\Omega^{(d-2)}$ is the metric on the unit
$(d-2)-$sphere, and $r$ is the invariant radius of the
$(d-2)-$sphere running through the point labeled by coordinates
$x,t$. We would like to consider generic charged  black holes, so
we will not restrict ourselves at this stage to any particular
gravity theory in $d$-dimensions. We assume only that the
spherically symmetric, vacuum sector has a Birkhoff-type theorem
which states that all such solutions are static (have a timelike
Killing vector) and can  be parametrized by two coordinate
invariant parameters, which we choose to be the mass, $M$ and
charge, $Q$. In this case there is always a coordinate system in
which, locally at least, the metric takes the form:
    \bea
    ds^2=-f(x;M,Q)dt^2 + {dx^2\over f(x;M,Q)}
    + r^2(x)d\Omega^{(d-2)}\, .
    \label{metric2}
    \eea
This `Schwarzschild-like' coordinate system is essentially unique.
The associated time coordinate $t$  we call the ``Schwarzschild
time'' for future reference. The corresponding functions
$f(x;M,Q)$ and $r(x)$ are essentially uniquely determined by the
dynamical equations of the particular theory under consideration.
We assume the existence of at least one
 event horizon, whose location $x_h(M,Q)$ is given
implicitly as a function of the mass and charge by:
    \bea
    f(x_h;M,Q)=0 \, .\label{horizon1}
    \eea
In the case that multiple horizons exist, $x_h$ will refer to the
outermost horizon. We stress that we do not need to assume any
particular form for the gravitational Lagrangian, only that a
Birkhoff-type theorem exists in the spherically symmetric sector.
However, as a specific example, we can again consider the
Reissner-Nordstr\"om solution, in which $r=x$ and $f=(1-2GM/r +
GQ^2/r^2)$, so that there are horizons at $r_\pm =
GM\pm\sqrt{G^2M^2-GQ^2}$. In this case $r_h=r_+$.

We now examine the thermodynamic behavior of generic charged black holes.
The following derivation of the black hole temperature is most
useful for the subsequent analysis. The basic
idea is to Euclideanize the solution (\ref{metric2}) by defining $t_E=-it$,
and
requiring the resulting solution exterior to the horizon to be regular. To
this end, we define new (Euclidean) coordinates:
    \bea
    R^2(x)&=& a^2 f(x;M,Q)\, ,\nonumber\\
    \alpha&=& t_E/a         \, ,      \label{euclid coords}
    \eea
where  the constant $a$ will determine the temperature of the
black hole. Note that the horizon is located at $R=0$. In these
coordinates, the radial part of the  metric reads:
    \bea
    ds^2_E=R^2d\alpha^2 +
    N(R) dR^2  \, .\label{euclid metric}
    \eea
This geometry will be regular for $R\geq0$, free from conical
singularities, providing that $\alpha$ is an angular coordinate,
whose period we assume to be $2\pi$, and the function $N(R)$ goes
to unity at $R=0$. These conditions determine the constant $a$ to
be:
    \bea a(M,Q)= {1\over f'(x_h;M,Q)} \, , \label{a} \eea where the
prime denotes differentiation with respect to $x$, and hence
requires the Euclidean time coordinate to be periodic, with range:
\bea 0\leq t_E \leq 4\pi/f'(x_h;M,Q) \, .\label{period} \eea In the
imaginary time formulation of finite temperature quantum field
theory the  periodicity of the Euclidean time is proportional to
inverse temperature. Applying this principle to the present
calculation
 yields the correct
 Hawking temperature:
\bea
T_H(M,Q)= {\hbar f'(x_h;M,Q)\over 4\pi} \, .
\label{temperature}
\eea
This expression agrees in all known cases
with semi-classical quantum field theoretic calculations of the
temperature of the thermal radiation emitted by black holes. Note
that we wish to interpret \req{temperature} as the temperature of
an exterior horizon. It must therefore by non-negative.
For the Reissner-Nordstr\"om black hole this
gives the condition $GQ^2\leq M^2$. More generally, this condition
restricts the charge and mass to have values for which $T_H$ as given in
\req{temperature} is non-negative. It  will play an important role below.

Once the Hawking temperature is determined, it is straightforward to
deduce the expression for the Bekenstein-Hawking entropy of the black hole
$S_{BH}(M,Q)$, which obeys the generalized first law of thermodynamics:
\bea
\delta M = T_H(M,Q) \delta S_{BH}(M,Q) + \Phi(M,Q)\delta Q \, .
\label{first law}
\eea
The second term is the contribution to the energy from the work required to
insert charge $\delta Q$ into the black hole, where $\Phi(M,Q)$ is
the electrostatic potential at the horizon. (Strictly this assumes
that the electrostatic potential vanishes at infinity.)

For the Reissner-Nordstr\"om solution, the Hawking temperature
is $T_{BH}=\hbar(r_+-r_-)/4\pi r_+^2 $, while the electrostatic
potential is $\Phi= Q/r_+$. It can easily be verified that these quantities
imply that the Bekenstein Hawking entropy as given in \req{rn entropy}
satisfies the generalized first law \req{first law}.

Recall that we are assuming that the theory under consideration, whatever it
is, admits a Birkhoff theorem, i.e. that  $M$ and $Q$
are the only
 diffeomorphism invariant parameters in the solution space. In this case
it is straightforward to deduce the general  form of the
reduced action that describes the dynamics of the spherically
symmetric sector of  an
isolated, generic charged black hole:
    \bea
    I^{red}= \int dt \left(P_M \dot{M} +
    P_Q \dot{Q} - H(M,Q)\right) \, ,       \label{reduced action}
    \eea
 where $P_M$ and $P_Q$ are the conjugates to $M$ and $Q$, respectively.
 The specific form of the Hamiltonian is irrelevant, except that it is
independent of $P_M$ and $P_Q$. This guarantees that $M$ and $Q$ are
constants of motion.

Although \req{reduced action} is motivated by completely general arguments, it
is reassuring that we can arrive at it in a more standard way. Start with
Einstein-Maxwell action in $d-$dimensions
    \bea
    I = - \int d^{d}x {\sqrt -g^{(d)}}
    \left[ \frac{R^{(d)}}{16\pi G_{d}} +
    {F_{A B}F^{AB}\over 4} \right]  \, ,\label{einstein maxwell}
    \eea
where $A,B=0,...,d-1$ and $F_{AB}=\partial_AA_B-\partial_BA_B$ is
a $d$-dimensional abelian field strength. If one substitutes into
\req{einstein maxwell} the generic spherically symmetric  form of
the metric (\ref{metric1}) and restricts the vector potential $A$
to be spherically symmetric as well, one obtains a  dimensionally
reduced `dilaton gravity' action in two spacetime dimensions, the
generic form of which is:
    \bea
    I^{eff}= {1\over 2 G} \int d^2 x
    \left [ D(r) R(g)+U(r)|\nabla r|^2 + V(r) -{1\over 4} W(r)
    F^{\mu\nu}F_{\mu\nu}\right]       \, ,  \label{dilaton action}
    \eea
where $\mu,\nu=0,1$. In this context $r(x,t)$ plays the role of a
dilaton field, and $D(r)$, $U(r)$, $V(r)$ and $W(r)$ are arbitrary
functions of $r$. Moreover, $F_{\mu\nu}=\partial_{[\mu} A_{\nu]}$
is the field strength associated with the spherically symmetric
 components of the
gauge potential $A_\mu$. For future reference we note that under a
gauge transformation $A_\mu\to A_\mu+\partial_\mu \lambda$
where $\lambda(x,t)$ is an arbitrary function of $x,t$. The boundary
values of the gauge function $\lambda$ will play an important role in
the subsequent analysis. For specific choices of the functions $D$, $U$, $V$
and $W$ this
action correctly describes the dynamics of the spherically symmetric
sector of a very large class of higher dimensional black holes. For
details see \cite{dk} and \cite{mk}.

The most general solution to the equations derived from
\req{dilaton action} does have a timelike Killing vector and  is
of the form (\ref{metric2}).  The details of the Hamiltonian
analysis can be found in \cite{mk}. Here we will simply summarize
the results. One proceeds as usual by parametrizing the metric:
    \bea
    ds^2=e^{2\rho}\left[-\mu^2 dt^2
    +(dx+\nu dt)^2\right]     \, ,    \label{adm metric}
    \eea
where $\mu$ and $\nu$ are the lapse and shift functions, which
play the role of Lagrange multipliers enforcing the two
constraints associated with diffeomorphism invariance in two
spacetime dimensions. $A_0$ is also a Lagrange multiplier that
enforces the Gauss law constraint. The only physical fields at
this stage are therefore the spatial metric $g_{11}= e^{2\rho}$,
the dilaton field, $r$, and the spatial component $A_1$ of the
vector potential. As with all diffeomorphism invariant theories,
the canonical Hamiltonian of the gravitational sector includes a
linear combination of constraints. In two spacetime dimensions,
this is also true of the electromagnetic sector. The canonical
Hamiltonian is therefore of the  form:
    \bea H_c = \int dx
    \left[ \nu {\cal F}
    + {\mu\over 2G} {\cal G} +
    A_0 {\cal J} \right] + H_{ADM}, \label{canonical hamiltonian}
    \eea
where $H_{ADM}$ is the ADM surface integral yielding the
Hamiltonian in the reduced action \req{reduced action}, ${\cal F}$ and
${\cal J}$ generate spatial diffeomorphisms and U(1) gauge
transformations, respectively, while ${\cal G}$ is the Hamiltonian
constraint generating time reparametrizations. All three are
functions of $\rho$, $r$, $A_1$ and their canonically conjugate
momenta. Once the constraints are imposed, the Hamiltonian reduces
to the surface term $H_{ADM}$, which depends on only two gauge
invariant parameters, namely the mass $M$ and charge $Q$. Finally,
we arrive at the reduced action \req{reduced action}.

\begin{figure}
\centerline{ \epsfxsize 4in
\epsfbox{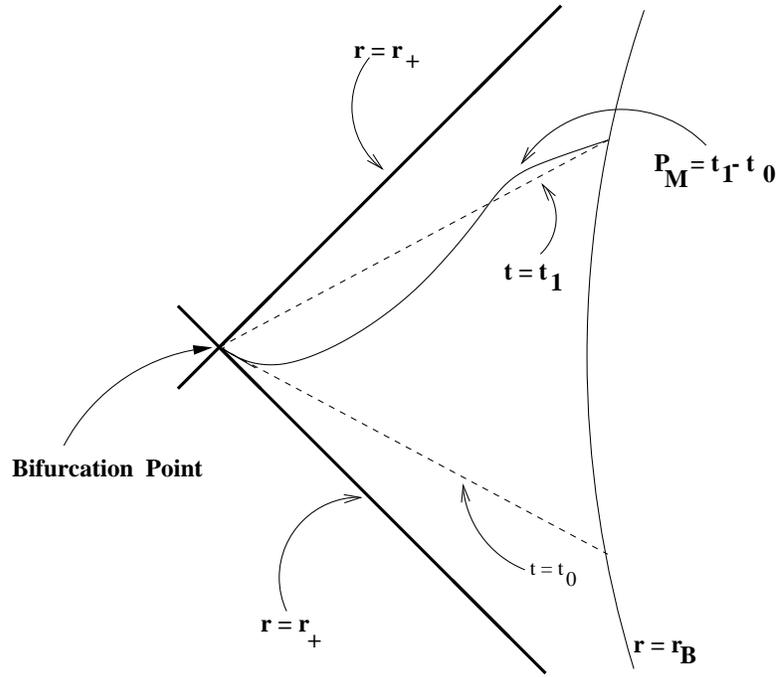}
}
\caption{\sl Kruskal diagram of exterior region of eternal 
Reissner-Nordstr\"om-like black hole with bifurcation horizon at 
$r=r_+$. Dashed lines denote surfaces of constant Schwarzschild time $t$. 
A slice of given $P_M=t_1 - t_0$ has a fixed difference between the 
Schwarzschild time $t$, which it approaches at infinity (or at radius 
$r_B$), and the Schwarzschild time $t_0$, which it approaches at the 
bifurcation point.}
\label{fig:kruskal}
\end{figure}

The specific form of $H_{ADM}$ depends on the boundary conditions that
are imposed. The choice of boundary conditions is in turn dictated by
the physical  circumstances. For the present purposes, the most useful
boundary conditions are those first considered by Louko and Whiting
\cite{lw} in the context of Schwarzschild black holes, and then later
generalized to  both uncharged \cite{shelemy}
and charged \cite{mk} black holes in generic two-dimensional dilaton
gravity. We consider eternal black holes with at least one bifurcative
horizon, so that there is an exterior region whose Kruskal diagram
is as shown in Fig. [1]. 
Note that Fig.[1] shows only the exterior wedge of the Kruskal diagram,
since that is all we need to consider for our analysis. 
In order to analyze the thermodynamic behaviour
of the black holes, we would like to consider solutions that can be analytically continued to the regular Euclidean solutions described above. We therefore
restrict to spatial slices that lie entirely in this exterior wedge. In particular, we follow Louko and Whiting\cite{lw} and require the
left hand side of the slice to approach the bifurcation point along a static
Schwarzshild slice, while the right side of the slice ends on a line 
of constant $r=r_B$ (i.e. a box of fixed radius $r_B$). Sample slices are shown in
Fig.[1]. These boundary conditions yield a boundary term in the Hamiltonian that depends on the mass and charge on the black hole, and on several
external variables, such as the radius of the box, the value of the electromagnetic gauge potential at the box. (For details of the boundary conditions, see \cite{mk}.) Moreover, on analytic continuation to periodic Euclidean time, the 
exterior wedge in Fig.[1] is mapped onto a ``cigar tube''. The closed end
of the cigar tube corresponds to the bifurcative horizon, and is regular
 providing the period of the Euclidean time coordinate is
chosen as described above.

Note that we started with a phase space consisting of three fields
and their conjugates. Since the Hamiltonian contains three first
class constraints that each generate a local gauge transformation,
the standard Dirac Hamiltonian analysis leads us to conclude that
there are no field theoretic physical modes in this class of
theories.   This is consistent with our assumption that the theory
admits a generalized Birkhoff theorem: the only diffeomorphism
invariant parameters in the solution space are the mass and
charge. It also agrees with our intuition about spherically
symmetric gravity and electromagnetism: there is no monopole
radiation in either theory.

The two diffeomorphism invariant observables $M$ and $Q$ can be
written explicitly as functions of the phase space variables. When
the constraints are satisfied, these functions are spatially
constant, and by virtue of Hamilton's equations they are also time
independent. Their canonical conjugates $P_M$ and $P_Q$ are
functionals of the phase space observables: i.e. integrals over
spatial slices. These phase space observables are not invariant
under arbitrary diffeomorphisms and gauge transformations. Their
invariance would contradict the generalized Birkhoff theorem which
allows only two invariant observables. However, $P_M$ and $P_Q$
are physical observables in the Hamiltonian sense because they are
invariant under local diffeomorphisms and gauge transformations
that vanish on the boundaries of the system. As shown in
\cite{kuchar},\cite{thiemann} for spherically symmetric gravity and in
\cite{gkl}
for the generic theory, for the  assumed boundary conditions in which
the metric approaches its Schwarzschild form at either end of the spatial 
slice, the momentum $P_M$ conjugate to $M$ is
proportional to the difference between the Schwarzschild times at
either end of the slice (see Fig.[1]). It is therefore invariant only under
local diffeomorphism, i.e. those that vanish at the boundaries of
the system. Similarly, the momentum $P_Q$ conjugate to $Q$ is not
invariant under all gauge transformations, only those that vanish
on the boundaries. Specifically, explicit calculation shows that
the following relationship holds between $P_Q$, the momentum
conjugate to $Q$ and $P_M$\cite{dk}:
\bea \delta P_Q =- \Phi \delta P_M +
\delta \lambda \, ,\label{momentum relation} \eea where the variations
refer to variations under a change in boundary conditions, $\Phi$
is the electrostatic potential at the boundary under
consideration, and $\delta \lambda$ is the variation in $U(1)$
gauge transformation $\lambda$ at the boundary. This relationship
will be important in what follows.

\section{Quantization: Toy Models}
We now have some information about the classical reduced
 phase space of black holes. However, an action of the form
(\ref{reduced action}), given in terms of constants of motion,
is not a traditional starting point for quantization, and it is in
general difficult to know how to proceed. Before considering the
physical system at hand, namely black holes, we will therefore look at a
couple of standard toy models that illustrate the utility, and validity of
our method.
\subsection{The Simple Harmonic Oscillator}

Consider the following action:
\bea
I= \int dt (P_M \dot{M}- M)
\label{reduced}
\eea
where $M\geq0$. Clearly this system is analogous to (\ref{reduced action}).
$M$ is a constant of motion, whose classical values is bounded below by
zero. The equations of motion imply that $P_M=t+constant$ is the
time variable in the problem.
It is possible to proceed once the boundary conditions on the
conjugate momentum $P_M$ are known. In particular, let us suppose that the
dynamics for this system is known to be periodic, with period $2\pi\over \omega_0$, for some angular frequency $\omega_0$. One could try to construct
self adjoint operators for $P_M$ and $M$ encorporating these boundary conditions, but it is significantly easier to first transform at the classical level
to variables in which the global structure of the phase space is easier to
deal with. Therefore consider the following transformation:
\bea
X&=&\sqrt{B(M)}\cos(P_M\omega_0)\\
P_X&=&\sqrt{B(M)}\sin(P_M\omega_0)
\label{sho_1}
\eea
In terms of the new variables $(X,P_X)$, the periodicity of $P_M$ is
manifest.
This transformation is canonical if and only if
$B(M)= {2M/\omega_0}+a_0$, where $a_0$ is an arbitrary constant. If
 we choose $a_0=0$, then the transformation is well-defined for all
$M\geq0$ as required. In this case, we have succeeded in mapping the
phase space $M,P_M$, which has the topology of a half cylinder (since
$M\geq0$ and $P_M$ is periodic, 
to the space $X,P_X$ which has the topology of the complete plane $R^2$, 
providing we include the origin, which corresponds
to $M=0$. Quantization is now straightforward. In particular note that
in terms of the new variables:
\bea
M = {\omega_0\over 2} (X^2 +P_X^2)
\eea
so that $M$ is effectively the energy of a harmonic oscillator, with
unit mass, and fundamental frequency $\omega_0$. By choosing the
usual measure and factor ordering one obtains the harmonic oscillator
spectrum for $M$:
\bea
M_n = \hbar \omega_0 (n+1/2)
\eea
This is of course no surprise since the boundary conditions that we
imposed on $M$ and $P_M$ were precisely those of the harmonic oscillator.
$B(M)= 2M/\omega_0$ in our canonical transformation is the action variable
associated with the angular coordinate $\alpha = \omega P_M$, and as
expected by the Bohr-Sommerfeld relation, it has an equally spaced spectrum.
What this example is meant to illustrate is that, starting from
variables that corresponded to a constant of motion and its canonical
conjugate, the boundary/periodicity conditions lead, via the canonical
transformation, to the correct harmonic oscillator spectrum. If this procedure
were only valid for the harmonic oscillator, it would not be of any use,
so we illustrate its utility in one more toy example before
going on to the black hole case.
\subsection{Bouncing Ball}
We now consider precisely the same action (\ref{reduced}), again with
the condition $M\geq0$, but with a different periodicity condition
on its conjugate. Suppose that we know physical solutions to be
periodic in $P_M$ with period $\Delta P_M = L\sqrt{2m\over M}$. We again
try a canonical transformation that accurately reflects this periodicity:
\bea
X&=&\sqrt{B(M)}\cos\left({2\pi P_M \sqrt{M}\over L \sqrt{2m} }\right)\\
P_X&=&\sqrt{B(M)}\sin\left({2\pi P_M \sqrt{M}\over L \sqrt{2m} }\right)
\label{bouncing}
\eea
In this case, the condition that the transformation be canonical is
satisfied if $B(M) = {2L\over \pi}\sqrt{2Mm}$. In terms of the new
variables, $B(M)$ again looks like the Hamiltonian for a harmonic
oscillator (note however that $B(M)$ is {\bf not} the Hamiltonican for
the physical system). Its spectrum, up to possible factor ordering
ambiguities is therefore:
\bea
B_n=\hbar (n+1/2)
\eea
The corresponding spectrum for the Hamiltonian, $M$ is:
\bea
M_n = {1\over 2m} \left[{\hbar\pi \over 2L} (n+1/2)\right]^2
\label{bouncing ham}
\eea
The physical interpretation of this example can be made explicit by
going to new coordinates $(q,p)$ defined by
\bea
M&=& {p^2\over 2m}\\
P_M&=& {m q\over p}
\eea
We therefore see that $M$ and $P_M$ are the Hamiltonian and time
corresponding to a free particle with position $q$ and momentum $p$.
The periodicity given above is that associated with a free particle
``bouncing'' between two walls (infinite potential barriers) a distance
$L$ apart \cite{kibble}.
The motion repeats when $q$ goes through $2L$, which, using
the above transformations gives the correct periodicity in terms of
the time variable $P_M$. Again we find that $B(M)= (2L/\pi)|p|$ is the
adiabatic invariant for this system. The  spectrum we obtained for
$|p|$, namely $|p|_n=\hbar\pi(n+1/2)/2L$, $n=0,1,2..$ is not quite the same
as that derived in most text books on quantum mechanics. The spectrum
derived by more traditional methods is:
\bea
|p|_n= {\hbar\pi n\over 2L} \quad\quad , \quad n=1,2,...
\eea
This difference between these two spectra is simply one of factor
ordering: our method can be made to yield the standard spectrum by
choosing a different factor ordering for the operators $X$ and $P_X$
in $B$. This example shows again that by knowing the
range/periodicity of the variables in (\ref{reduced}) it is possible
to deduce the spectrum of the corresponding
adiabatic invariant and energy, at least
up to factor ordering.
\section{Black Hole Quantization}
As shown by the examples above, the first step towards quantization
in terms of the present variables is to determine the ranges  of the various
phase space observables. It is reasonable to keep $M$
non-negative, while $Q$ must be a real number satisfying the
 condition $T_H(M,Q) \geq 0$. It can be shown that
 this condition can generically be expressed as a bound
on the entropy in terms of the charge, namely:
    \beq
    S_{BH}(M,Q)\geq S_0(Q) \, , \label{entropy minimum}
    \eeq
where the function $S_0(Q)$ depends on the theory under
consideration. For example, it can be verified by examining
\req{rn entropy} that in the case of Reissner-Nordstrom black
holes the condition, $G^2M^2\geq GQ^2$
requires \req{entropy minimum} with $S_0(Q) = \pi Q^2/\hbar$. For
dimensionally reduced Einstein-Maxwell theory in $d$ dimensions
with $S^{(d-2)}$ spherical symmetry, one has
    \bea
    S_0(Q)= K_{(d)}Q^{(d-2)/(d-3)} \, , \label{G(Q)}
    \eea
where
    \bea K_{(d)} = (1/4) (A_{d-2}/G_{d})^{(d-4)/2(d-3)}
    (8\pi/(d-2)(d-3))^{(d-2)/2(d-3)} \, ,\label{Kd} \eea
and $A_{d-2}=2\pi^{(d-1)/2}/\Gamma((d-1)/2))$ is the area of the
unit $d-2$ sphere. It is interesting to note that in all cases
except $d=4$, the entropy bound depends explicitly on the
gravitational constant $G_d$.

As discussed in Section 2, in black hole geometrodynamics \cite{kuchar}
$P_M$ gives the difference of the Schwarzschild times at the ends
of the spacelike slice running across the Kruskal diagram. When
analytically continued to the Euclidean spacetime, this variable
becomes imaginary and, as motivated from semi-classical
thermodynamics, periodic, with period given by the inverse Hawking
temperature (\ref{period}). We will therefore henceforth make the
assumption that we must identify:
    \bea P_M
    \sim P_M+{1\over
    T_H(M,Q)} \, . \label{PM period}
    \eea

We are now in a situation familiar in classical mechanics, where
there exists a periodic, angular variable. Akin to the
action-angle formulation of the harmonic oscillator, we `unwrap'
our gravitational  phase space, by transforming to a set of
unrestricted variables. Consider the following transformation
$(M,Q,P_M,P_Q) \rightarrow (X,Q,\P_X,\P_Q)$:
    \bea
    &&X=\sqrt{\hbar B(M,Q)\over \pi}
    \cos(2\pi P_M T_H(M,Q)/\hbar) \, ,\nonumber\\
    &&\Pi_X= \sqrt{\hbar B(M,Q)\over \pi }
    \sin (2\pi P_M T_H(M,Q)/\hbar)    \, ,        \label{grav}\\
    &&Q=Q \, ,\nonumber\\
    &&\P_Q=\Pi_Q(M,P_M,Q,P_Q) \, ,                 \label{charge}
    \eea
where $B(M,Q)$ and $\Pi_Q(M,P_M,Q,P_Q)$ are functions that will be determined
by the
condition that the transformation be canonical.
Transformations \req{grav} yield a pair of non-periodic variables in a
way that  encorporates directly the correct periodicity
of $P_M$.

A straightforward calculation reveals that, up to a total
variation in independent variables $M$ and $Q$:
    \bea
    \Pi_X\delta X+\Pi_Q \delta Q=
    P_M \left(T_H{\partial B\over\partial M}\right)
    \delta M+\left(\Pi_Q+ P_M T_H {\partial B\over\partial Q}\right)
    \delta Q, \label{canonical conditions}
    \eea
so that this transformation is canonical when
    \bea
    {\partial B\over\partial M}
    &=&{1\over T_H(M,Q) }\, ,\\
    P_Q &=&\Pi_Q+
    P_M T_H {\partial B\over\partial Q}\, .  \label{canonical 1}
    \eea
{}From the first law \req{first law} we see that
\beq
{\partial B\over \partial M} = {\partial S_{BH}\over \partial M}
\eeq
It therefore follows that
    \bea
    B(M,Q)= S_{BH}(M,Q)+F(Q) \, ,\label{B1}
    \eea
where $S_{BH}$ is the Bekenstein-Hawking entropy associated with a
black hole of mass $M$ and charge $Q$, while $F(Q)$ is an
arbitrary function of the charge. This function can be fixed by
noting that the transformation (\req{grav}) maps the fundamental
domain of the initial phase space variables $(M,P_M)$, $0\leq 2\pi
P_M T_H(M,Q)/\hbar <2\pi$, to the exterior of a disc of
finite radius, such that $B(M,Q)\geq S_0(Q)+F(Q)$,
    \bea
    B(M,Q) = {2\pi\over\hbar} \left({1\over 2}X^2
    + {1\over 2}\P_X^2\right). \label{B2}
    \eea
Here $S_0(Q)$ is the function that determines the minimum entropy
in terms of charge for the generic theory. To avoid ambiguity in
quantization caused by the necessity of imposing boundary
conditions at the minimal radius of $S_{BH}(M,Q)=S_0(Q)$, it is
natural to remove this round ``hole'' in phase space plane. We,
therefore, demand that $F(Q)= -S_0(Q)$. This achieves
two crucial things: first that the
phase space topology in the new variables is trivial -- a {\em complete}
two-dimensional plane -- and secondly that
\req{entropy minimum} is identically satisfied. Moreover, the
precise extremal limit corresponds to the origin of this plane.
With this choice,
$\Pi_Q$ is uniquely determined to be:
    \bea
    \Pi_Q = {\hbar\over e}\chi
    + {\hbar\over 2\pi}S'_0(Q)\alpha, \label{canonical 2}
    \eea
where $'$ denotes differentiation with respect to $Q$ and we have
defined the  variables: $\chi = {e\over \hbar}( P_Q+\Phi
P_M)$ and $\alpha = 2\pi P_M T_H(M,Q)$.

An important reservation regarding the canonical transformations
of the above type is that they are performed for the Euclidean
theory in which only the periodicity of the variable $\alpha=2\pi
P_M T_M$ makes sense. The possibility of such canonical
transformations and subsequent quantization altogether can be
called in question because of their Euclidean status. The
justification of this procedure, however, follows from the
important fact that the Euclideanization of the canonical action
\req{reduced action} is not just the analytic continuation to the
imaginary range of the time $t$, but {\em simultaneously} the same
continuation for the momenta variables. In contrast to the usual
situation this leaves us with the same (up to an overall
$i$-factor) {\em real} canonical action in Euclidean variables
(for the usual Wick rotation procedure the kinetic term of the
canonical action acquires an extra imaginary unit factor). This
special type of Euclideanization is not universal, because it is
possible only for cyclic momenta not entering the Hamiltonian as
in Eq. \req{reduced action}. This explains why the canonical
equations of motion retain the same form in Euclidean regime as in
the Lorentzian one and why they can be rewritten in terms of the
same Poisson bracket $\{M,P_M\}=1$ (under the usual Wick rotation
such a bracket becomes imaginary for Euclidean variables). The
further quantization as a promotion of ($M,P_M$) (or canonically
related $(X,\Pi_X)$) to the operator level subject to canonical
commutation relations, $[X,\Pi_X]=i\{X,\Pi_X\}=i$, is
straightforward and runs as follows.

{}From \req{B1} and \req{B2} we see that in the gravitational sector
of $(X,\Pi_X)$ for fixed charge $Q$
    \bea
    S_{BH}-S_0(Q) = {2 \pi\over \hbar}
    \left({X^2\over2} +
    {\P_X^2\over 2}\right). \label{harmonic oscillator 1}
    \eea
This operator is the Hamiltonian of a harmonic
oscillator with the mass and frequency both equal to $\hbar/2\pi$
. Since the domain of variables $X$ and $\Pi_X$ is an entire
two-dimensional plane, their quantization becomes trivial. To be
precise, with zero boundary conditions at infinity of this plane
one can define self-adjoint operators in the space of
square-integrable wavefunctions and thus obtain the quantum
mechanical spectrum for the entropy of generic charged black
holes:
    \bea
    S_{BH}= 2\pi \left(n+{1\over 2}\right)+S_0(Q)
    \qquad n = 0,1,2,... \label{entropy spectrum}
    \eea
It is important to note
that due to vacuum fluctuations the limit of extremal black
hole, $S_{BH}=S_0(Q)$, corresponding classically to the origin of the
phase-space
plane $(X,\Pi_X)$, cannot be
achieved at the quantum level. \req{entropy spectrum} also
implies that near extremal black holes,
despite being potentially  macroscopic objects (i.e. large $Q$, and
horizon area) are necessarily highly quantum objects\footnote{We are grateful
  to Valeri Frolov for raising this issue.} in the sense of
corresponding to small quantum number, $n$.

To complete the analysis we now need to quantize the
electromagnetic sector and derive the spectrum for the operator
$Q$. This requires knowledge of the boundary conditions on its
conjugate $\Pi_Q$.
For compact gauge group $U(1)$ \req{momentum relation}
suggests that the linear combination $\chi=e\lambda/\hbar=e( P_Q+
\Phi P_M)/\hbar$ is an ``angular coordinate'' with period $2\pi$,
where $e$ is the electromagnetic coupling. In particular, suppose that
there exists a
charged scalar field $\psi$ minimally coupled to the vector
potential via the covariant derivative
\bea
D_\mu\psi=(\partial_\mu -
i {e\over \hbar} A_\mu)\psi \, ,
\label{covariant derivative}
\eea
where $e$ gives the strength of the electromagnetic coupling.
Under a gauge transformation, $A_\mu\to A_\mu
+\partial_\mu \lambda$, $D_\mu$ is invariant providing that
$\psi \to e^{(ie \lambda/\hbar)}\psi$. Thus, $\lambda$ has the
period claimed.
Examining \req{canonical 2} we now see that $\Pi_Q$ is a function of
two angular coordinates $\chi$ and $\alpha$
 which, according to
arguments given above are both periodic with period $2\pi$
    \bea
    \chi\sim\chi+2\pi n_1,\,\,\,
    \alpha\sim\alpha+2\pi n_2.
    \eea
 It is therefore
necessary to identify the phase space points
    \bea (Q, \Pi_Q) \sim
    (Q, \Pi_Q + 2\pi n_1 {\hbar\over e}
    +  n_2 \hbar S'_0(Q)).                    \label{identify}
    \eea

In the coordinate representation, ${\hat Q}=-i\hbar
\partial / \partial \Pi_Q$, the wave functions for
charge eigenstates take the form
    \bea \psi_Q(\Pi_Q) =
    {\rm (const)}\times \exp(i Q \Pi_Q/\hbar).    \label{charge eigenstates}
    \eea
This wave function is single valued under the identification
\req{identify} when for any integer $n_1$ and $n_2$ the following
combination
    \bea
    n_1 {Q\over e} +
    n_2 {Q\over 2\pi} S_0'(Q)= n_3    \label{phase 1}
    \eea
is also given by an integer number $n_3$. This immediately results
in two quantization conditions with two integer numbers $m$ and
$p$
    \bea
    &&{Q\over e}=m,                        \label{Q-spectrum}\\
    &&{Q\over 2\pi} S_0'(Q)=p,              \label{e-spectrum}
    \eea
which together imply not only the black hole charge $Q$ is
integer multiple of the $U(1)$ charge $e$, but also that the latter is
subject to a quantization condition-- the allowed value of $e$ is
constrained in terms of  $m$ and $p$.

In order to determine the implications of these quantization
conditions one must know the specific form of $S_0(Q)$. For
concreteness, take first Reissner-Nordstr\"om black holes with
$S_0(Q)= \pi Q^2/\hbar$.
The charged black hole itself is characterized by two integer
numbers $n$ and $p$ which determine its horizon area (entropy) and
charge
    \bea
    &&S_{BH}=2\pi n+\pi(p+1),  \label{entropy}\\
    &&Q^2=\hbar p.              \label{charge1}
    \eea
The quantum number $p$ determines the charge of the
quantum black and hence its  minimal entropy $S_0=\pi(p+1)$.
The quantum number $n$ determines the excited level of the black
hole over the "vacuum", $n=0$ for which the entropy achieves its
minimum value. Classically this vacuum would correspond
to an extremal black hole with minimal admissible value of its mass
$M=Q/\sqrt G$ and entropy $S_0(Q)$ for a given charge $Q$. It is
a remarkable feature of our analysis that this
classical lower bound on the mass is never actually saturated due
to vacuum fluctuations of the horizon -- the $+\pi$ contribution
in \req{entropy} survives in the critical limit $n=0$ of a
charged quantum black hole. Thus, extremal black holes are not in the physical
spectrum (at least for our Weyl type
quantization).

 Finally, there exists a third
quantum number $m$ which shows that the  charge $Q$ is multiple of
the $U(1)$ coupling constant $e$. However, this coupling
constant is not completely arbitrary,
 since \req{Q-spectrum} and
\req{e-spectrum} fix the value of fine structure constant in terms
of integer numbers $m$ and $p$:
    \bea
    {e^2\over\hbar}={p\over m^2}   \, .   \label{fsqc}
    \eea
Thus, $e^2/\hbar$ must be a rational number

We would like to stress that the essential qualitative features of the
spectrum are generic to all
spherically symmetric black holes, and not specific to the
Reissner-Nordstrom
case.
For completeness, we consider spherically symmetric black
holes in $d$ dimensions, for which $S_0(Q)$ is given by
\req{G(Q)}. In this case, the condition \req{e-spectrum}
generalizes to
    \bea
    {e^2\over\hbar}=\frac{p^{2(d-2)/(d-3)}}{m^2}(d-3)^2
    \left(\frac{8\pi\hbar G_d}
    {A_{d-2}}\frac{d-3}{d-2}\right)^{(d-4)/(d-2)}.
    \eea
Interestingly, the fine structure constant spectrum for
dimensionalities other than four depends on the d-dimensional
gravitational coupling constant $G_d$.

\section{Adiabatic Invariants and the Black Hole Emission Spectrum}

With the above transformations, one can prove
Bekenstein's conjecture that the horizon area is an adiabatic
invariant \cite{bm}. We can express
(\ref{harmonic oscillator 1}) in terms of the area variable as follows:
    \begin{equation}
    A - 4G\hbar S_0(Q) =
    {8\pi G} \left(\frac{X^2}{2} +
    \frac{\P_X^2}{2} \right). \label{area1}
    \end{equation}
Now, consider the integral
$$ {\cal J}_X= \oint \P_X~dX $$
where the integration is over one complete period and ${\cal J}_X$
is the {\it angle variable} for the oscillator in the action-angle
formalism. Now, it is well known that for a periodic system, under
an adiabatic (slow) perturbation with a time dependent parameter
$\l(t)$, satisfying the condition $\d\l/\l \ll \d t/T$ for a time
interval $\d t$ ($T$ is the time period of oscillation), ${\cal
J}_X$ remains invariant, although both the energy and the
fundamental frequency can change considerably over a period of
time \cite{landau}. For the harmonic oscillator Hamiltonian under
consideration, the above integral is simply the area of a circle
of radius squared $(A-4G\hbar S_0(Q))/4G$. Thus, it follows that
    \begin{equation}
    {\cal J}_X  =  \pi \frac{A-4G~\hbar S_0(Q)}
    {4G}
    \end{equation}
is an adiabatic invariant. In addition, note that the assumed periodicity of
the phase space involving $Q$ and $\P_Q$ implies that
    \begin{equation}
    {\cal J}_Q= \oint \Pi_Q dQ
    \end{equation}
is an adiabatic invariant as well. Consequently, from the expression for
$\Pi_Q$ in (\ref{canonical 2}), it follows that the $U(1)$ charge $Q$ (and
thus
any function of $Q$) is an adiabatic invariant as well. Thus, we conclude that
the area observable is itself an adiabatic invariant, thus proving the
conjecture made in \cite{bm} that the horizon
area is an adiabatic invariant in quantum gravity.
The effect upon quantization becomes evident as well. According to
the Bohr-Sommerfeld quantization rules, action variables are always
quantized \cite{pauling}.
Thus the area operator of quantum gravity must be quantized
as seen in the preceding calculations. Note however, that
the appearance of the `ground state entropy' of $\p$ could not have been
guessed from old quantum theory, without solving the Schr\"odinger
equation.

Exponentiating the entropy, we obtain the actual degeneracy of the
black hole in the level $n$: \bea g(n) = e^{2\pi (n+1/2) + S_0(Q)}
\eea It is interesting to note that the since $g(0) \neq 1$, the
ground state is degenerate. The fact that $g(0)$ is not an integer
should not be considered too disturbing at this stage. We have
taken the semi-classical expression for the Hawking temperature to
be exact when determining the periodicity of the Euclidean time
coordinate. In this context it is amusing and perhaps relevant
 that $g(0)=23$ (i.e. is an integer)
if one simply replaces $T_{H}$ by $0.998 T_{H}$ in
the canonical transformations \req{grav}, and everywhere in the
subsequent analysis.

Now let us consider a physical process in which the black hole emits
a photon by making a quantum jump from one level to the next lower
level. The exact mechanism of such a jump is irrelevant in the
current analysis.
To calculate the frequency of the emitted photon, we go
back to the entropy formula (\ref{bh entropy})
 applied to a Reissner-Nordstr\"om black hole (\ref{rn entropy}).
Assuming that the black hole decays
by emitting just one photon with the lowest allowed frequency
$\o_0$ (for simplicity we assume
uncharged particle emission and four dimensions),
its initial and final masses are
$M + \hbar \o_0$ and $M$ respectively, and the
following relation holds:
$$ S(M+ \hbar \o_0,Q) - S(M,Q) = S(n+1) - S(n) = \p $$
from where it follows that
\bea
\o_0 = \frac{ (r_+ - r_-) \p}{A}\, .
\label{freq}
\eea

This frequency for the $Q\rightarrow 0$ limit agrees
with that found in \cite{bm} up to factors of order unity.
However, unlike there, here the fundamental frequency can be made
arbitrarily small by going near the extremal limit (i.e. small
quantum number $n$), while keeping the charge, and mass large.
So, although the emission spectrum is in general
discrete, in the near extremal limit, the Hawking spectrum will become
almost continuous, as predicted by semi-classical analyses and
also by the area spectrum of loop quantum gravity \cite{rovelli}.

Finally, our quantization procedure indicates that at the end stage of
radiation form the black hole, there is a Planck size
remnant which is left behind, corresponding to the `zero point area'
which inevitably results from the uncertainty principle. It is
tempting to speculate that this remnant will contain any information
that may have entered the black hole before and during its evaporation
process. It would be an interesting exercise to construct an explicit
model of radiation by introducing an interaction Hamiltonian and
compute the transition amplitudes between the levels.

\section{Summary and Conclusions}
In this paper we have quantized the charged black hole sector of
gravity, and derived both the area (entropy) and the charge
spectrum by first deriving the reduced action for  the spherically
symmetric sector of generic charged black holes, and then
providing input about the black hole thermodynamics by assuming
periodicity of the phase space variable conjugate to the black
hole mass. Although the periodicity is motivated by the
Euclideanized black hole solutions, so that its relevance to
Minkowskian black holes is not beyond doubt, it is not altogether
unnatural to associate a fundamental time scale with physical
black holes (see \cite{louko}, \cite{kastrup} and \cite{kastrup_strobl}).
Moreover, the beauty and simplicity of the resulting analysis, as
well as its remarkable generality, suggests that it must have
something to do with the real world at the quantum level.

We close by attempting to interpret some of the more intriguing
implications of the spectrum that we have derived. The
interpretation of the first quantization condition
\req{Q-spectrum} is obvious -- the black hole can absorb and emit
integer number of particles with fundamental charge $e$. The
second condition \req{fsqc} implies that the value of this
fundamental constant must be related to the  integer parameters of
the black hole $m$ and $p$. This can be interpreted in one of two
ways. If one considers the black hole states to be fundamental,
then the presence of a charged black hole in the universe would
fix the value of the fine structure constant and hence the
fundamental unit of electric charge. This is somewhat analogous
to the way the presence of a single Dirac magnetic monopole of
magnetic charge $g$ requires all electric charges in its vicinity
to be quantized in units of $2\pi\hbar/g$. This is also
reminiscent of Coleman's old idea \cite{bigfix}, that wormhole
physics may fix the conventional fundamental constants of nature.

It would seem that the big fix mechanism of \cite{bigfix} based
upon sharply peaked distribution functions of Euclidean quantum
gravity is conceptually different from our mechanism of single
valued wavefunctions in the space of periodic variables. However,
under a closer examination these concepts might turn out more
closely related than one could have anticipated. Indeed, we do not
know yet the correct interpretation of the black hole
thermodynamics input -- the periodicity in Euclidean time with the
inverse Hawking temperature period. The formalism of Euclidean
quantum field theory, as is well known, can originate from two
distinctively different physical situations -- from the
description of thermodynamical ensemble (statistical, i.e. not
pure, state) or from the description of classically forbidden
transitions between pure states -- quantum mechanical underbarrier
tunneling. Quite amazingly, in quantum gravity these two functions
of the Euclidean formalism are not yet clearly separated. That is
why this field of science was designed to have a special name --
Euclidean quantum gravity -- very ambiguous and flexible for
possible interpretations. Indeed, the Euclidean section of the
Schwarzschild solution can, on one hand, be regarded as a saddle
point of the path integral for the statistical partition function
and, on the other hand, can be viewed as a classical configuration
interpolating in the imaginary time between the two causally
disconnected spacetime domains -- right and left wedges of the
Kruskal diagram. Our requirement of periodicity in the imaginary
time apparently can be viewed as a kind of consistency of quantum
states in these two domains, or the finiteness of the
semiclassical underbarrier transition amplitude between them
(remember that the Hawking periodicity requirement is based on the
absence of conical singularity which is, in its turn, motivated by
the regularity of the semiclassical distribution). So the
amplitudes not satisfying this requirement can be regarded as
suppressed to zero, just like in the Coleman big fix paradigm, the
vanishing probability of having fundamental constants that violate
the superselection rules imposed by coherent states of baby
universes. In this respect, our derivation of restrictions on the
fine structure constant can be regarded as a big fix mechanism
revisited in context of black hole physics.

The physical significance of our result and its foundation of the
above type is still not clear, and at this stage one might raise a
number of objections to this conclusion. In particular, one might
ask how this picture stands the coexistence of different black
hole with different charges each demanding its own fundamental
value of $e$, one might call in question the very interpretation
of the restriction on $e$, which could be related not to the
quantization consistency requirement, but rather indicate to the
structure of interferometric patterns caused by the black hole
scattering of quantum electrons.

Alternatively,  one can take the viewpoint that  the charged fields
are
fundamental, so that our quantum black hole states must somehow be
created out of these fundamental fields. In this case,  the condition
\req{fsqc} simply means that the two charge quantum numbers $m$
and $p$ are not independent. However, \req{fsqc} only yields
consistent solutions for $m$  and $p$ if
$e^2/\hbar$ is a rational number. So we are still in the position that
the black hole quantum mechanics places a non-trivial condition
on the fine structure constant.

Clearly,
the actual experimental value of the fine structure constant, known
with incredibly high accuracy to be $4\pi\hbar/e^2=137.03608...$, cannot
be approximated by the rational number \req{fsqc} with
reasonably small integers $p$ and $m$. So whatever viewpoint is adopted,
this mechanism remains
very speculative. It does, however, deserve further study, not least
because of the fascinating consequence that  a fundamental constant of
nature is restricted by the consistency of the black hole quantum
mechanics.

After the original version of this paper was written we learned
about two other derivations of spectra for charged black holes
\cite{VW} and \cite{MRLP}. Both the principles of derivation and
the quantitative results of these works differ essentially from
our approach. These two works bear in common the fact that within
the full or partial Hamiltonian reduction one of the variables --
describing the proper time in a spacelike slicing of the
Reissner-Nordstrom or Kerr-Newman geometry between the two
horizons at $r_\pm$ -- has a finite range related to the mass of
black hole. The periodic extension of this Lorentzian time
variable from this range yields a quantization condition similar
to the one proposed in \cite{kastrup}. It should be emphasised,
however, that these works suffer either from a rather wishful
handling of the Wheeler-DeWitt equation \cite{VW} or from a very
restrictive type of spacetime foliation that does not cover the
entire spacelike infinity \cite{MRLP}. (The latter paper includes
angular momentum as well as charge, but the actual Hamiltonian
reduction is conjectured rather than explicitly performed).  The
different foundations of these methods from ours  result in
quantitatively different conclusions. Interestingly, the
quantities that acquire the equally spaced  spectra in \cite{VW},
\cite{MRLP} and in our case \req{entropy spectrum} are
respectively $S_+-S_-$, $S_+ +S_-$ and $S_+-S_0$, where
$S_\pm\equiv \pi r_\pm^2/G\hbar$  are the entropies associated
with the outer and inner horizons, while $S_0=\pi Q^2/\hbar$ is
the intermediate quantity -- the BPS bound \req{entropy minimum}.
It is important, however, that despite this qualitative
resemblance, references \cite{VW} and \cite{MRLP} do not
incorporate the second quantization condition (39) and therefore
do not predict restrictions on the fine structure constant.

The fact that these methods yield only qualitatively similar results to
ours is indicative of their conceptually different foundations.
 In our opinion, the assumption of
periodicity in Lorentzian time, whether it is introduced
by hand as in \cite{kastrup} or effectively induced by
spacetime foliations pinched between inner and outer horizons as
in \cite{VW,MRLP}, does not seem convincing. On the other hand, the
appeal to Euclidean quantum gravity in the form of the
underbarrier dynamics in imaginary time interpolating between
the wedges of the Kruskal diagram looks more promising.  However,
its ultimate justification might require
knowledge of the as yet unfinished chapter
of  Hawking's virtual black holes theory \cite{vbh}.

\vspace{1cm}
\noindent {\bf Acknowledgments}

\noindent  S.D. and G.K. would like to thank Viqar Husain for
helpful comments and encouragement. G.K. also thanks Valeri Frolov and
the gravity group at the Theoretical Physics Institute, University of
Alberta
for useful discussions.  A.B. is grateful to V.Rubakov
for helpful discussion. We also acknowledge the partial support of
the Natural Sciences and Engineering Research Council of Canada.
This work was also supported by the Russian Foundation for Basic
Research under the grant No 99-02-16122 and by the grant of
support of leading scientific schools No 00-15-96699.  This work
has also been supported in part by the Russian Research program
``Cosmomicrophysics''.

\end{document}